\documentclass[twocolumn,showpacs,preprintnumbers,amsmath,amssymb,superscriptaddress, prl,floatfix]{revtex4-1}
\usepackage{bm}
\usepackage{graphicx}
\usepackage{dcolumn}
\usepackage{tabularx}
\usepackage{bm}
\usepackage{epstopdf}
\usepackage{amsmath}
\usepackage{color}
\usepackage{here}

\begin{document}

\preprint{Preprint}

\title{Coherent and incoherent aspects of polariton dynamics in semiconductor microcavities}

\author{N. Takemura}
\email[E-mail: ]{naotomot@phys.ethz.ch}
\affiliation{Laboratory of Quantum Optoelectronics, Physics Institute, \'Ecole Polytechnique F\'ed\'erale de Lausanne, CH-1015, Lausanne, Switzerland}
\author{M. D. Anderson}
\affiliation{Laboratory of Quantum Optoelectronics, Physics Institute, \'Ecole Polytechnique F\'ed\'erale de Lausanne, CH-1015, Lausanne, Switzerland}
\author{S. Biswas}
\affiliation{Laboratory of Quantum Optoelectronics, Physics Institute, \'Ecole Polytechnique F\'ed\'erale de Lausanne, CH-1015, Lausanne, Switzerland}
\author{M. Navadeh-Toupchi}
\affiliation{Laboratory of Quantum Optoelectronics, Physics Institute, \'Ecole Polytechnique F\'ed\'erale de Lausanne, CH-1015, Lausanne, Switzerland}
\author{D. Y. Oberli}
\affiliation{Laboratory of Quantum Optoelectronics, Physics Institute, \'Ecole Polytechnique F\'ed\'erale de Lausanne, CH-1015, Lausanne, Switzerland}
\author{M. T. Portella-Oberli}
\affiliation{Laboratory of Quantum Optoelectronics, Physics Institute, \'Ecole Polytechnique F\'ed\'erale de Lausanne, CH-1015, Lausanne, Switzerland}
\author{B. Deveaud}
\affiliation{Laboratory of Quantum Optoelectronics, Physics Institute, \'Ecole Polytechnique F\'ed\'erale de Lausanne, CH-1015, Lausanne, Switzerland}

\date{\today}

\begin{abstract}
The interaction between coherent polaritons and incoherent excitons plays an important role in polariton physics. 
Using resonant pump-probe spectroscopy with selective excitation of single polariton branches, we investigate the different dephasing mechanisms responsible for generating a long-lived exciton reservoir. 
As expected, pumping the upper polariton results in a strong dephasing process that leads to the generation of a long lived reservoir.
Unexpectedly, we observe an efficient reservoir creation while exciting only the lower polariton branch when the detuning is increased towards positive detuning. 
We propose a simple theoretical model, the polaritonic Bloch equations, to describe the dynamics of the system.

\end{abstract}


\maketitle
\section{I. Introduction}
Exciton-polaritons are interesting quasi-particles emerging from the strong coupling between an exciton and a cavity photon, inheriting a vanishingly small effective mass from the photon while preserving exciton-mediated interactions. 
The long spacial and temporal coherence of polaritons enables a variety of interesting physical phenomena including: Bose-Einstein condensation \cite{Kasprzak2006,Christopoulos_PRL07}, superfluidity \cite{Amo2009,Amo2009b_Nature09,Kohnle2011, Kohnle2012}, and quantized vorticity\cite{Lagoudakis2008,Lagoudakis2009}. 
While many of these phenomena can be captured by the Gross-Pitaevskii equations, which are equations of motion of the coherent polariton wave-function, it is well known that incoherent excitons also play a crucial role in the modification of the temporal dynamics of polaritons. 
We have previously shown that when polariton branches at ${\bm k}\sim0$ are coherently excited by a spectrally broad laser pulse, the coherent exciton polarization converts into incoherent exciton population due to both pure and excitation induced dephasing (EID) \cite{Takemura2015b}, this process is best described using excitonic Bloch equations (EBEs)\cite{Rochat2000}. 
However, since the excitonic Bloch equations are written in the exciton-photon basis and not the natural basis of the system (\textit{i.e.,} polariton basis), they cannot fully capture some of the complex dynamics arising from the distinct polariton branches.
In this paper, we perform a more thorough study on polariton dephasing by exciting single polariton branches with a spectrally narrow pulse and propose a modified model, the polaritonic Bloch equations (PBEs), to introduce branch dependent dephasing. 
We find that only when the lower polariton branch is excited at negative cavity detunings, does the system approximately follow the coherent dynamics described by the Gross-Pitaevskii equations.
In all other cases, a strong dephasing (EID and pure dephasing) convert the polariton population into a long lived incoherent exciton population.
These results clearly define the regime in which coherent polaritonic devices can function.

\begin{figure}
\includegraphics[width=0.45\textwidth]{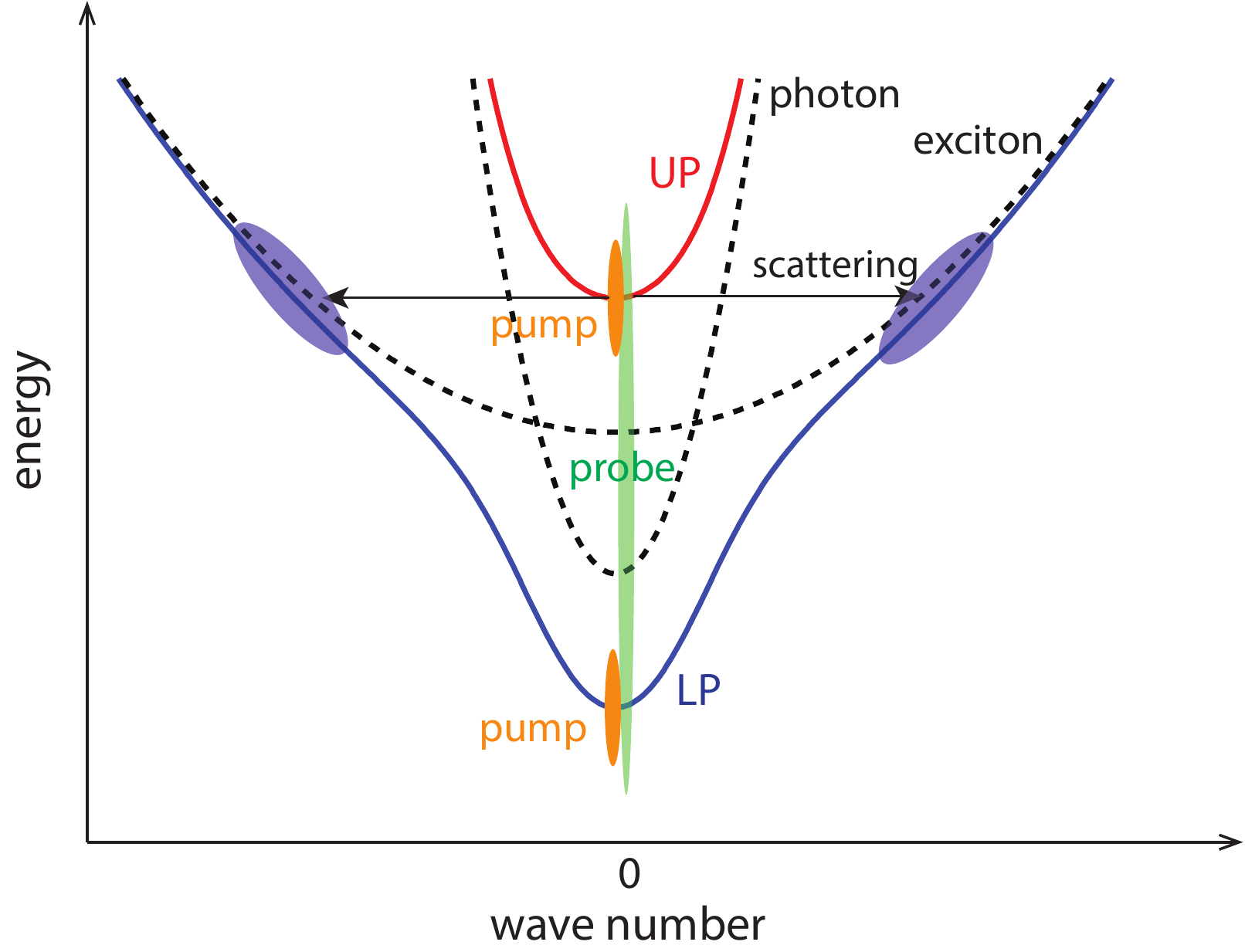}
\caption{(color online). Scheme of excitation configuration and possible scattering scenario in experiments. The lower (LP) and upper polariton (UP) energy-momentum dispersions at slight negative cavity detuning (-1.2 meV) are shown. The dashed black lines represent exciton and photon energy-momentum dispersion. At this cavity detuning, the scattering of excited upper polaritons into the large momentum exciton-like polariton state is allowed, while the scattering of the lower polaritons can be neglected.}
\label{fig:scheme}
\end{figure}

\section{II. Experiment}
The experiments are performed with a high quality GaAs-based microcavity \cite{Stanley1994} at the cryogenic temperature of 4K. 
The sample is a single 8 nm In$_{0.04}$Ga$_{0.96}$As quantum well sandwiched between two GaAs/AlAs distributed Bragg-reflectors (DBRs) with Rabi splitting energy $2\Omega$=3.45 meV at zero cavity detuning \cite{Stanley1994,Kohnle2011}. 
We use a two-beam pump-probe configuration with degenerate beams at $\bm k\sim 0$ $\mu$m$^{-1}$.
The sample is excited with a spectrally narrow pump pulse and probed with a spectrally broad probe pulse (See Fig. \ref{fig:scheme}). 
The optical pulses are generated spectrally broadband, a few hundred femtoseconds, by a Ti:Sapphire laser and then the pump pulse is spectrally narrowed (to $\sim$0.5 meV) using a single grating pulse shaper.
This configuration enables the excitation of a single polariton branch, either the lower polariton (LP) or the upper polariton (UP), while probing with a spectrally broad probe pulse centered between the lower and upper-polariton peaks. 
The transmitted probe beam is then detected using a heterodyne detection technique \cite{Takemura2014a}. 
In order to avoid the biexciton effects \cite{Takemura2014,Takemura2014a}, we employ co-circularly polarized pump and probe pulses. 

\begin{figure}
\includegraphics[width=0.45\textwidth]{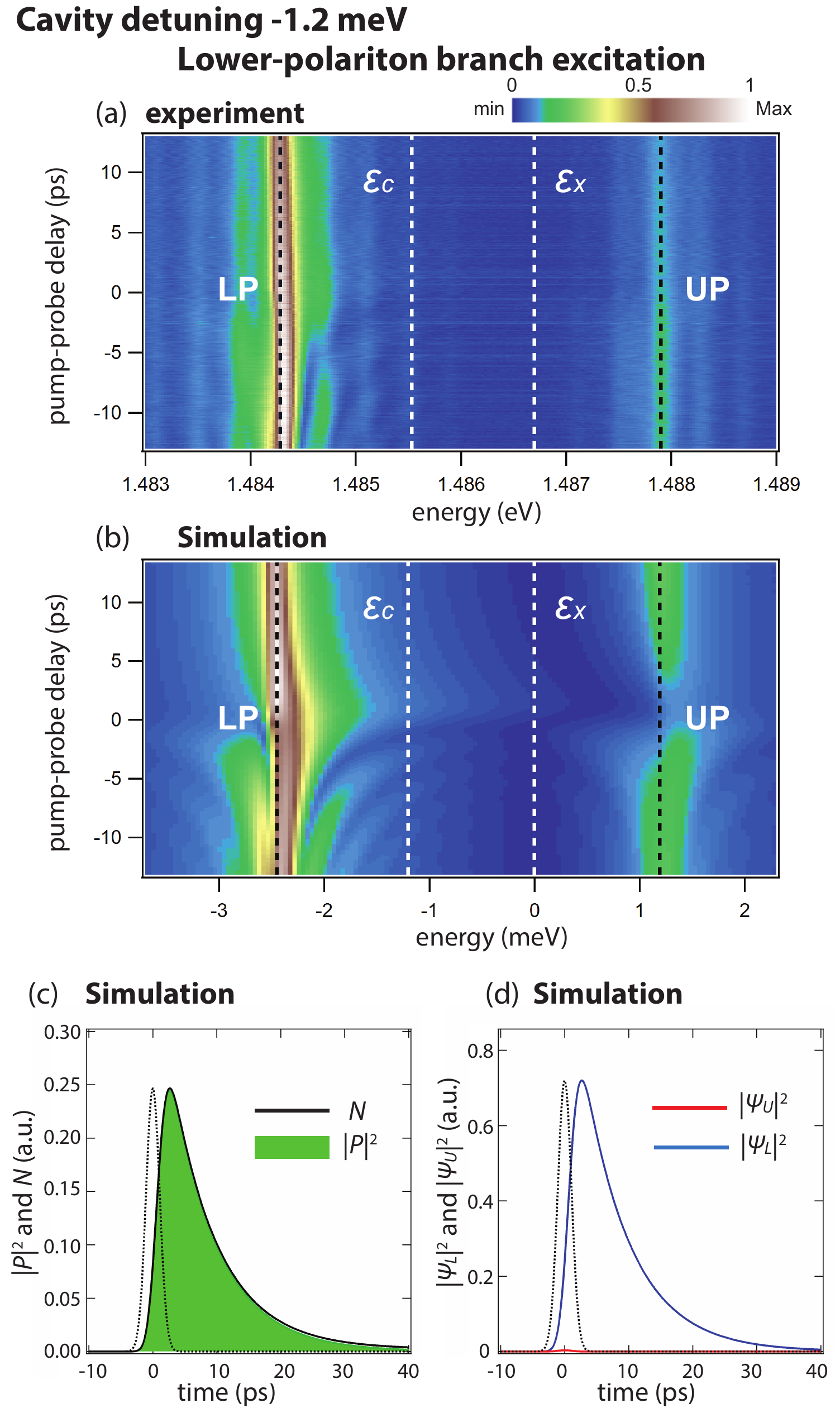}
\caption{(color online) Measured (a) and simulated (b) probe transmission are shown as a function of energy and time delay between pump and probe pulse. 
The pump pulse selectively excites only the lower polariton branch with an intensity of 7.4$\times$10$^{12}$ photons/pulse/cm$^{2}$ (500 $\mu$W). 
The black dashed lines represent the lower and upper-polariton peak energies without pump pulse. 
The white dashed lines are the cavity photon and exciton energies. 
The simulations of real time evolution of exciton population and polarization are represented in (c). 
The solid and filled lines are respectively the time evolution of population $N(t)$ and polarization $|P(t)|^2$. 
The simulated time evolutions of lower $|\psi_L|^2$ and upper polariton probabilities $|\psi_U|^2$ are presented in (d)}
\label{fig:exp-sim_LP}
\end{figure}

The experimentally obtained probe spectra as a function of pump-probe delay are shown for LP (Fig. \ref{fig:exp-sim_LP} (a)) and UP (Fig. \ref{fig:exp-sim_UP} (a)) excitation at a cavity detuning of -1.2 meV. 
In order to compare the two cases, we focus on the delay dependence of the energy shift of the probe at the lower polariton branch energy. 
When the pump pulse populates the LP branch (Fig. \ref{fig:exp-sim_LP}), a blue shift is observed for the LP branch which gradually decreases, reaching zero at positive delay.
Meanwhile, at negative delays, the LP energy shift is observed longer than -10 ps. 
In contrast, when the pump pulse populates the UP (Fig. \ref{fig:exp-sim_UP}), the energy shift of the LP branch of the probe presents an opposite behaviour, a long-lived energy shift is present at positive delays ($>$10 ps), while gradually decreasing at negative delays. 
Similar to the results with the spectrally broad pump \cite{Takemura2015b}, we can understand these phenomena by considering the presence of the long lived incoherent exciton population generated by polariton dephasing.
 
The two contrastive cases indicate that at the cavity detuning -1.2 meV, dephasing exists only when the UP branch is populated. 
Specifically, Fig. \ref{fig:exp-sim_LP} (a) is representative of the coherent limit while Fig. \ref{fig:exp-sim_UP} (a) shows the typical behaviour in the presence of interactions with an incoherent exciton population generated through polariton dephasing. 
Therefore, these results demand a model which can account for asymmetric dephasing rates between lower and upper polaritons, something not captured by the EBEs. 
For this purpose, we propose polaritonic Bloch equations (PBEs), which is the polaritonic basis version of the EBEs presented in Refs. \cite{Rochat2000,Takemura2015b}. 

\section{III. Theory}
In order to derive the PBEs, we start from the EBEs:
\begin{eqnarray}
i\hbar \dot{N}&=&-i\Gamma_x N-2i(\Omega-2g_{\rm pae}N){\rm Im}[PE^{*}]\nonumber\\
i\hbar \dot{P}&=&\tilde{\epsilon}_xP+\Omega E+g_0NP-2g_{\rm pae}NE\\
i\hbar \dot{E}&=&\tilde{\epsilon}_cE+\Omega P-g_{\rm pae}NP - f_{\rm ext},\nonumber
\label{eq:EBE}
\end{eqnarray}
where population $N$, polarization $P$, and electric field $E$ are given as $N({\bf x},t)=\langle\hat{\bm \psi}_{x}^{\dagger}\hat{\bm \psi}_{x}\rangle$, $P({\bf x},t)=\langle\hat{\bm \psi}_{x}\rangle$, and $E({\bf x},t)=\langle\hat{\bm \psi}_{c}\rangle$ with a boson exciton (photon) field operator $\hat{\bm \psi}_{x(c)}$. 
$\tilde{\epsilon}_{x}$ and $\tilde{\epsilon}_{c}$ are the generalized energies of exciton and photon including decay rates. 
Since the pump and probe beams are degenerate at ${\bm k}\sim 0$, we neglect photonic energy-momentum dispersion. 
These are defined as,
\begin{eqnarray}
\tilde{\epsilon}_{x}&=&\epsilon_x-i\gamma_x\nonumber\\
\tilde{\epsilon}_{c}&=&\epsilon_c-i\gamma_c
\end{eqnarray}
Here, $\epsilon_{x(c)}$ and $\gamma_{x(c)}$ are respectively the eigen energy and decay rate of exciton (photon). 
Using the decay rate of the exciton population $\Gamma_x$, the exciton dephasing $\gamma_x$ is defined as \cite{Shah1999,Baars2000,Ciuti1998},
\begin{equation}
\gamma_x=\Gamma_x/2.
\end{equation} 
Lastly, $g_0$ is the exciton-exciton interaction constant with energy shift. $g_{\rm pae}$ is the strength of photon-assisted exchange scattering. 

The main idea of EBEs is to factorize the expectation value $\langle\hat{\bm \psi}_{x}^{\dagger}\hat{\bm \psi}_{x}\hat{\bm \psi}_{x}\rangle$ in equations of motion as $\langle\hat{\bm \psi}_{x}^{\dagger}\hat{\bm \psi}_{x}\hat{\bm \psi}_{x}\rangle\simeq\langle\hat{\bm \psi}_{x}^{\dagger}\hat{\bm \psi}_{x}\rangle\langle\hat{\bm \psi}_{x}\rangle=N({\bf x},t)P({\bf x},t)$. This is contrast to the derivation of the Gross-Pitaevskii equations (GPEs), where we apply an approximation $\langle\hat{\bm \psi}_{x}^{\dagger}\hat{\bm \psi}_{x}\hat{\bm \psi}_{x}\rangle\simeq\langle\hat{\bm \psi}_{x}^{\dagger}\rangle\langle\hat{\bm \psi}_{x}\rangle\langle\hat{\bm \psi}_{x}\rangle=|P({\bf x},t)|^2P({\bf x},t)$ assuming a coherent state. The strength of the EBEs is to include both coherent $P({\bf x},t)$ and incoherent $N({\bf x},t)$ parts of the polariton dynamics, while GPEs assume that polaritons are fully coherent. There is a close analogy between the EBEs and optical Bloch equations (OBEs). In an analogy to OBEs, the decay rates of the population $\Gamma_x$ and polarization $\gamma_x$ can be regarded as the counterparts of $1/T_1$ and $1/T_2$

Now, using a matrix ${\bf M}$, we summarize the equations of motion of polarization and electric field parts as  
\begin{equation}
i\hbar\frac{d}{dt}\left( \begin{array}{c}
P \\
E \\
\end{array} \right)
={\bf M}
\left( \begin{array}{c}
P \\
E \\
\end{array} \right)
+\left( \begin{array}{c}
\tilde{g}NP-2g_{\rm pae}NE \\
-g_{\rm pae}NP-f_{\rm ext} \\
\end{array} \right),
\end{equation}
where the matrix ${\bf M}$ is given by
\begin{equation}
{\bf M}
=
\left( \begin{array}{c c}
\tilde{\epsilon}_{x} & \Omega\\
\Omega & \tilde{\epsilon}_{c}\\
\end{array} \right).
\end{equation}
The matrix {\bf M} representing the linear coupling between the polarization and electric field can be diagonalized by introducing new basis $\psi_{L}$ and $\psi_{U}$ defined as,
\begin{equation}
\left( \begin{array}{c}
\psi_L \\
\psi_U \\
\end{array} \right)
=
\left( \begin{array}{c c}
X & C\\
-C & X\\
\end{array} \right)
\left( \begin{array}{c}
P \\
E \\
\end{array} \right).
\label{eq:trans}
\end{equation}
$X$ and $C$ are generalized Hopfield coefficients defined as
\begin{eqnarray}
X=\sqrt{\frac{1}{2}\left(1+\frac{\tilde{\epsilon}_{c}-\tilde{\epsilon}_{x}}{\sqrt{(\tilde{\epsilon}_{c}-\tilde{\epsilon}_{x})^2+(2\Omega)^2}}\right)}
\end{eqnarray}
\begin{eqnarray}
C=-\sqrt{\frac{1}{2}\left(1-\frac{\tilde{\epsilon}_{c}-\tilde{\epsilon}_{x}}{\sqrt{(\tilde{\epsilon}_{c}-\tilde{\epsilon}_{x})^2+(2\Omega)^2}}\right)}
\end{eqnarray}
Due to the presence of the imaginary part of $\tilde{\epsilon}_{x}$ and $\tilde{\epsilon}_{c}$ (decay rates), the Matrix ${\bf M}$ is non-hermitian and as a consequence, the transformation of Eq. \ref{eq:trans} is non-unitary. 
In fact, both $X$ and $C$ are also complex values. 
Finally, with a direct calculation, the EBEs in Eq. 1 are rewritten as,
\begin{eqnarray}
i\hbar \dot{N}&=&-i\Gamma_x N-2i\left(\Omega-2g_{\rm pae}N\right)\nonumber\\
& &\cdot\left(|X|^2{\rm Im}[\psi_L\psi_U^{*}]-|C|^2{\rm Im}[\psi_U\psi_L^{*}]\right.\\
& &\left.+|\psi_L |^2{\rm Im}[XC^*]-|\psi_U |^2{\rm Im}[CX^*]\right)\nonumber\\
i\hbar \dot{\psi_L}&=&(\tilde{\epsilon}_L+g_{LL}N)\psi_L+g_{LU}N\psi_U-Cf_{\rm ext}\\
i\hbar \dot{\psi_U}&=&(\tilde{\epsilon}_U+g_{UU}N)\psi_U+g_{UL}N\psi_L-Xf_{\rm ext}.
\label{eq:PBE}
\end{eqnarray}
\begin{figure}
\includegraphics[width=0.45\textwidth]{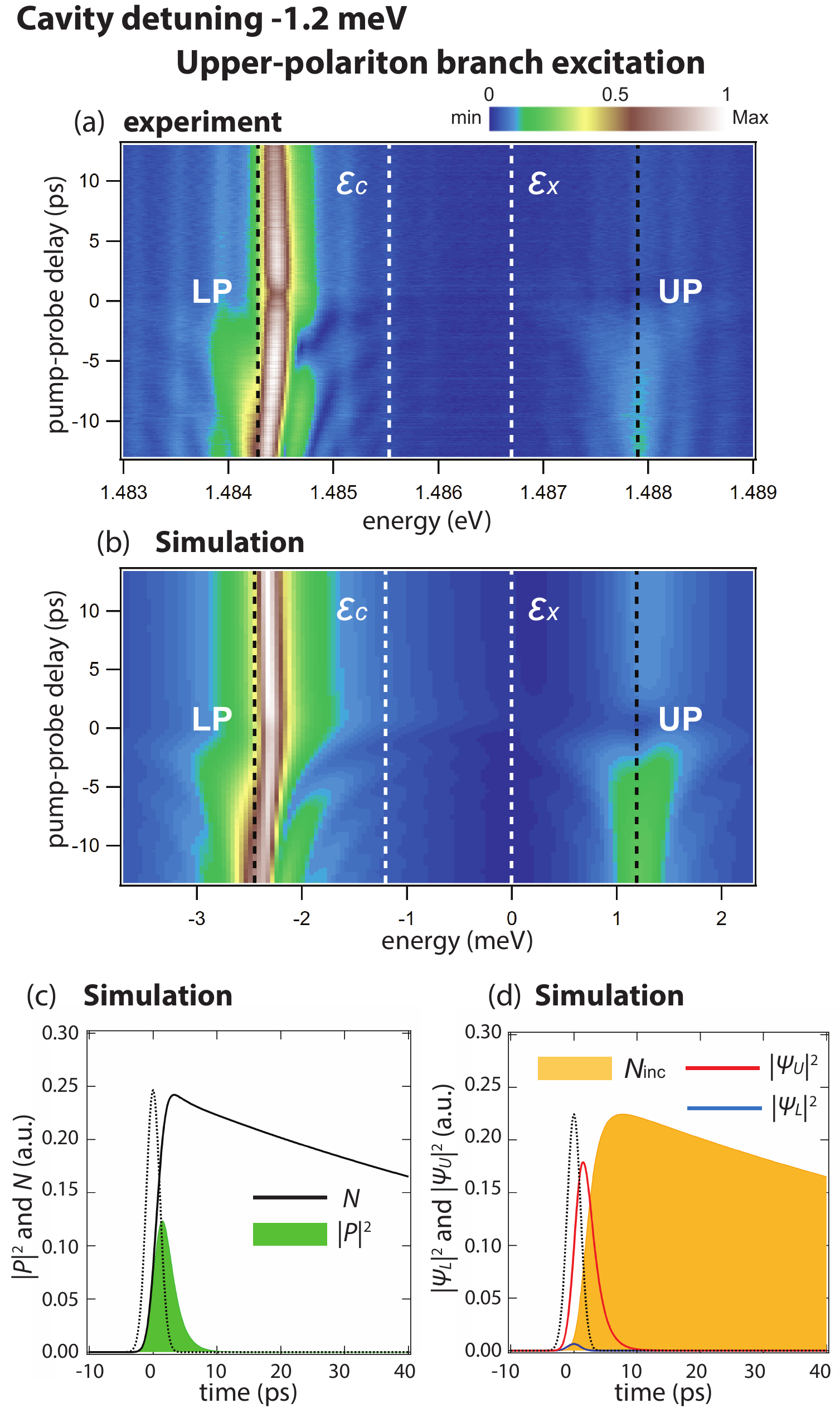}
\caption{(color online) Measured (a) and simulated (b) probe transmission are shown as a function of energy and time delay between pump and probe pulse. 
The pump pulse selectively excites only the upper polariton branch with an intensity of 7.4$\times$10$^{12}$ photons/pulse/cm$^{2}$ (500 $\mu$W). 
The black dashed lines represent the lower and upper-polariton peak energies without pump pulse. 
The white dashed lines are the cavity photon and exciton energies. 
The simulations of real time evolution of exciton population and polarization are represented in (c). 
The solid and filled lines are respectively the time evolution of population $N(t)$ and polarization $|P(t)|^2$, with the excitation pulse indicated by a black dotted line. 
The simulated time evolutions of lower $|\psi_L|^2$, upper polariton wave probabilities $|\psi_U|^2$, and incoherent exciton population $N_{\rm inc}$ are presented in (d)}
\label{fig:exp-sim_UP}
\end{figure}\begin{figure*}
\includegraphics[width=0.9\textwidth]{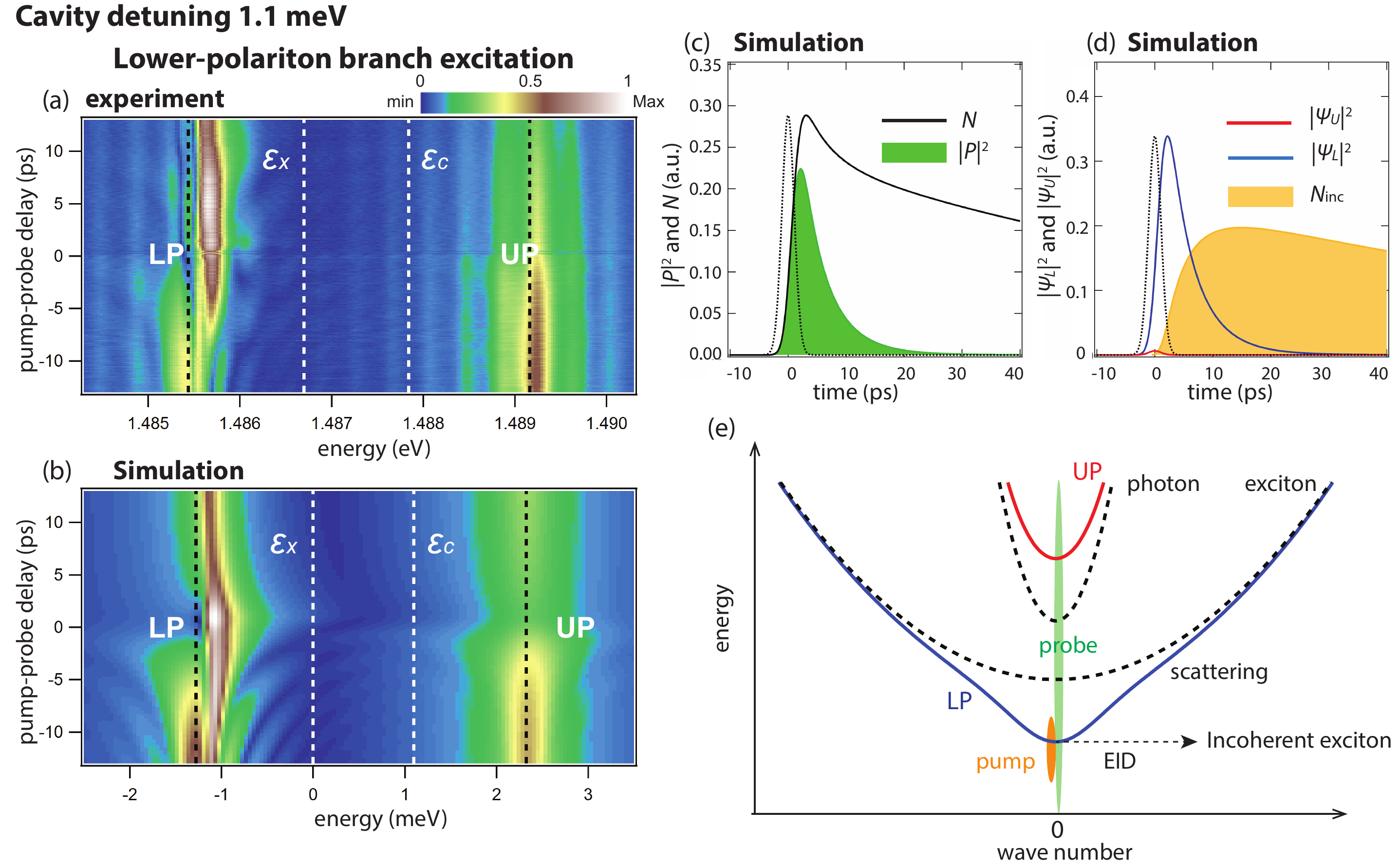}
\caption{(color online) Measured (a) and simulated (b) probe transmission are shown as a function of energy and time delay between pump and probe pulse for the cavity detuning at 1.1 meV. 
The pump pulse selectively excites only the lower polariton branch with an intensity of 7.4$\times$10$^{12}$ photons/pulse/cm$^{2}$ (500 $\mu$W). 
The black dashed lines represent the lower and upper-polariton peak energies without pump pulses. 
The white dashed lies are the cavity photon and exciton energies. 
The simulations of real time evolution of exciton population and polarization are represented in (c). 
The solid and filled lines are respectively the time evolution of population $N(t)$ and polarization $|P(t)|^2$. 
The simulated time evolutions of lower $|\psi_L|^2$, upper polariton probabilities $|\psi_U|^2$, and incoherent exciton population $N_{\rm inc}$ are presented in (d). (e): Scheme of excitation configuration. The experiment suggests the presence of the incoherent generation process through EID at 1.1 meV cavity detuning even though the scattering into the reservoir with a large wave-vector is prohibited.}
\label{fig:scheme_pos}
\end{figure*}
The above set of equations represent the newly proposed model which, because the new basis $\psi_L$ and $\psi_U$ can be interpreted as the lower and upper polariton basis, we name the Polaritonic Bloch Equations. 
$\tilde{\epsilon}_L$ and $\tilde{\epsilon}_U$ are complex eigen energies of lower and upper polaritons given by 
\begin{eqnarray}
\tilde{\epsilon}_L&=&\frac{1}{2}\left(\tilde{\epsilon}_{x}+\tilde{\epsilon}_{c}-\sqrt{(\tilde{\epsilon}_{c}-\tilde{\epsilon}_{x})^2+(2\Omega)^2}\right)\\
\tilde{\epsilon}_{U}&=&\frac{1}{2}\left(\tilde{\epsilon}_{c}+\tilde{\epsilon}_{x}+\sqrt{(\tilde{\epsilon}_{c}-\tilde{\epsilon}_{x})^2+(2\Omega)^2}\right).
\end{eqnarray}
Moreover, $g_{LL}$ and $g_{UU}$ are complex interaction constants that contain energy renormalization (real part) associated with $g_0$ and $g_{\rm pae}$ and EID (imaginary part) through $g'$. 
On the other hand, $g_{LU}$ and $g_{UL}$ represent a coupling (real part) and population transfer (imaginary part) between the lower and upper polaritons. 
They are written as,
\begin{eqnarray*}
g_{LL}&=&X^2\tilde{g}-3XCg_{pae}\\
g_{UU}&=&X^2\tilde{g}+3XCg_{pae}\\
g_{LU}&=&-XC\tilde{g}+(C^2-2X^2)g_{pae}\\
g_{UL}&=&-XC\tilde{g}+(2C^2-X^2)g_{pae}.
\end{eqnarray*}
The underlying idea of the PBEs is that the polaritonic wave-functions $\psi_L$ and $\psi_U$ are coupled to the exciton population $N$. 
The advantage of the polaritonic basis compared to the exciton-photon basis is that both the pure dephasing and EID can be introduced only in a single polariton branch as is experimentally observed.
For example, we can introduce dephasing only in the upper polariton by introducing additional imaginary parts to $\tilde{\epsilon}_{U}$ and $g_{UU}$ such that,
\begin{eqnarray}
\tilde{\epsilon}_{U}&\rightarrow& \tilde{\epsilon}_{U}-i\gamma^*_{UU}\\
g_{UU}&\rightarrow& g_{UU}-ig'_{UU}. 
\end{eqnarray}
It is worth noting that $N$ is the total exciton population that includes both coherent and incoherent parts. 
The incoherent component of the exciton population can be defined as,
\begin{equation}
N_{\rm inc}({\bf x},t)=N-|P|^2=N-|X\psi_L-C\psi_U|^2.
\end{equation} 
For the numerical calculation of pump-probe spectra, we employ the coupled-mode approximation in the same way as in Ref. \cite{Takemura2015b}. 
The pump (probe) pulse is introduced as a Gaussian pulse, 
\begin{eqnarray}
f^{pu(pr)}_{\rm ext}&=&F^{pu(pr)}\exp\left(-\frac{(t-t_{pu(pr)})^2}{2\tau_{pu(pr)}^2}\right)\nonumber\\
& &\cdot\exp\left(-i\omega^{pu(pr)}(t-t_{pu(pr)})\right).
\end{eqnarray}
The pulse durations of the pump $\tau_{pu}$ and probe $\tau_{pr}$ are set respectively as 1.5 and 0.35 ps. 
The ratio of the squared amplitudes of the pump ($|F^{pu}|^2$) to the probe ($|F^{pr}|^2$) is set to be 4 and $g_{0}\cdot |F^{pu}|^{2}=0.12$ (meV)$^{3}$. 
We display numerically calculated pump-probe spectra for the lower and upper polariton excitation cases in Fig. \ref{fig:exp-sim_LP} (b) and \ref{fig:exp-sim_UP} (b), respectively. 
For these simulations, $\Gamma_x$ and $\gamma_c$ are set to be 0.01 meV and 0.1 meV respectively \cite{Takemura2015b}. 
Since we introduce dephasing only for the upper polariton, we set $\gamma_x^*=0$ and $g'=0$. 
The exciton interaction constants are same as in Ref. \cite{Takemura2015b}: $g_{\rm pae}=0.3g_0$. 
The pure dephasing and EID strength for the upper polariton are $\gamma^*_{UU}=0.1$ meV and $g'_{UU}=0.8g_0$. 

\section{IV. Results}

At first, we consider the case when the only the LP branch is excited.
It can be seen in Fig. \ref{fig:exp-sim_LP} (a),(b), that both experiment and simulation show a decrease of the lower polariton energy shift of the pump-probe spectrum that is faster in the positive delay than in the negative one. 
The real time evolution of the exciton population $N(t)$ and the square of the polarization $|P(t)|^2$ for LP excitation are shown in Fig. \ref{fig:exp-sim_LP} (c). 
This clearly shows that the exciton population $N(t)$ can be replaced by the square of the polarization $|P(t)|^2$, \textit{i.e.,} $N(t)=|P(t)|^2$, which is a signature of the coherent limit. Furthermore, the simulated real time evolution of the lower $|\psi_L(t)|^2$ and upper polaritons $|\psi_U(t)|^2$ (Fig. \ref{fig:exp-sim_LP} (d)) indicates that the polarization can be approximated by the lower polariton wave-function, $|P|^2=|X\psi_L-C\psi_U|^2\simeq|X|^2|\psi_L|^2$. 
Considering these approximations $N\simeq|X|^2|\psi_L|^2$, the dynamics of the lower polariton is reduced to the conventional Gross-Pitaevskii equation. 
\begin{equation}
i\hbar \dot{\psi_L}=(\tilde{\epsilon}_L+g_{LL}|X|^2|\psi_L|^2)\psi_L-Cf_{\rm ext}
\end{equation}

Second, we will consider the case when only the UP branch is excited.
Here, the pure dephasing $\gamma^*_{UU}$ and large EID $g'_{UU}$ of the UP quickly convert a large portion of exciton polarization into an incoherent exciton population $N_{\rm inc}$ (Fig. \ref{fig:exp-sim_UP} (c),(d)).
The strong broadening of the upper polariton at zero delay in the pump-probe spectrum, shown in Fig. \ref{fig:exp-sim_UP} (a), is the evidence for the onset of dephasing. 
Since the exciton population has a long life time ($\sim \hbar/\Gamma_x$), the lower polariton branch of the probe spectrum presents a long lived energy blue shift at positive delays.

\begin{figure}
\includegraphics[width=0.45\textwidth]{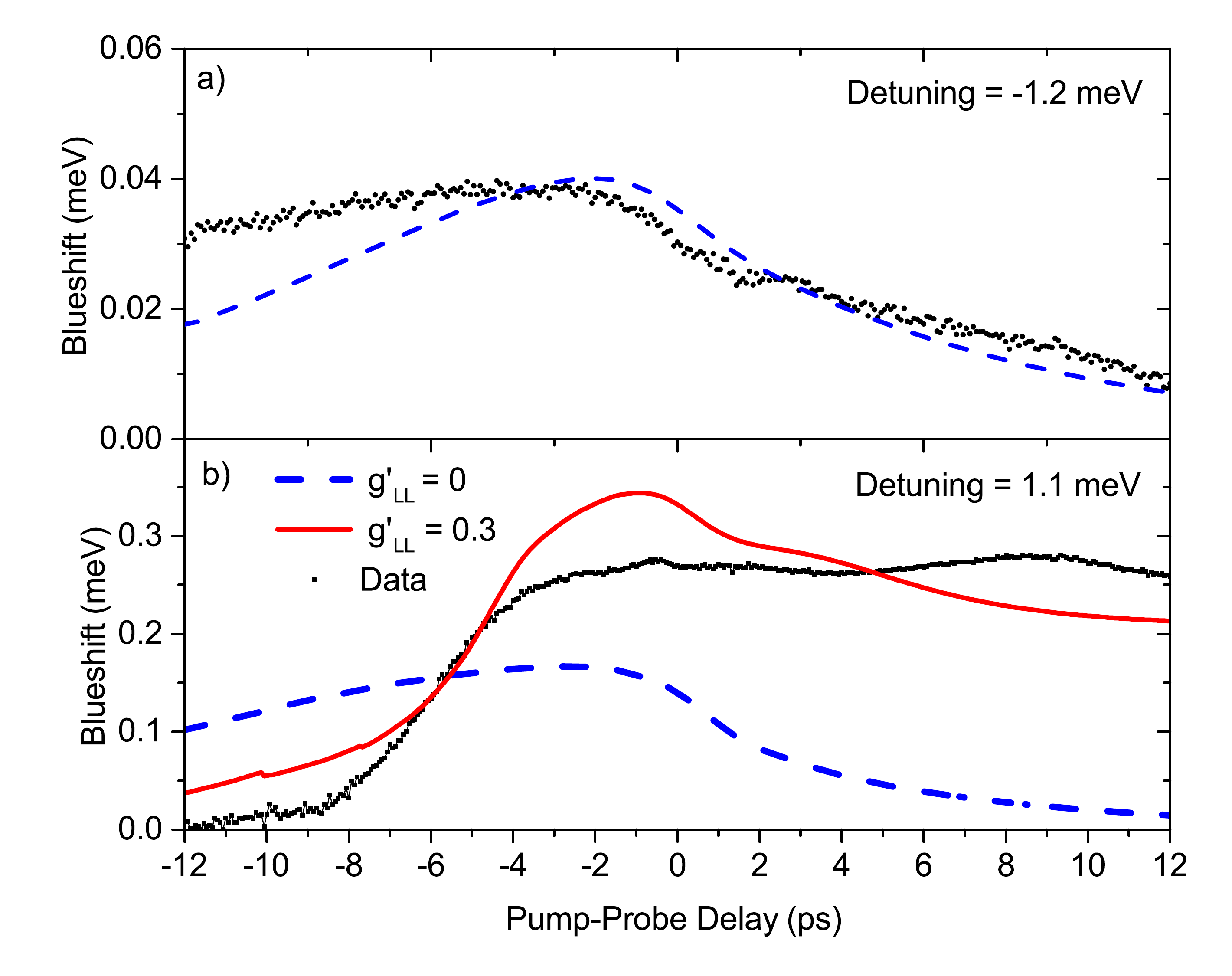}
\caption{(color online) Comparison of the extracted blueshift of the LP for a) negative detuning and b) positive detuning. Optimal parameters with no LP EID match well the experimentally observed blueshift relaxation for the negative detuning. For the positive detuning an EID of 0.3 was required to reproduce the long lived blue shift.)}
\label{fig:bsfit}
\end{figure}

Now, let us discuss the case of positive cavity detuning at 1.1 meV (Fig. \ref{fig:scheme_pos}). 
As expected, when the UP branch is excited, a strong dephasing occurs in the same way as in the negative detuning (not shown). 
What is more surprising is that when only the LP branch is excited (Fig. \ref{fig:scheme_pos} (a)), the energy blue shift of the lower polariton branch does not decrease at positive delays like it does for negative cavity detuning case (see Fig. \ref{fig:exp-sim_LP} (a)). 
In order to reproduce this behaviour, we need to include a finite EID for the lower polariton branch,
\begin{eqnarray}
g_{LL}&\rightarrow& g_{LL}-ig'_{LL}. 
\end{eqnarray}    
For the numerical simulation in Fig \ref{fig:scheme_pos} (b), we set $g'_{LL}=0.3g_0$. 
The other parameters are same as in the negative cavity detuning case. 
Although the upper polariton population is not involved in the process, the EID of the lower polariton converts the coherent exciton fraction of the lower polariton into a long lived incoherent exciton population (See Fig. \ref{fig:scheme_pos} (d)). 
The extracted blue shift for both positive and negative detunings is shown in Fig. \ref{fig:bsfit}, in the case of positive detuning the addition of $g'_{LL}$ is required to better match the data.
Now, in an analogy to the Gross-Pitaevskii equations coupled to a reservoir \cite{Carusotto2013,Nardin2009,Lagoudakis2011,Manni2012}, it is tempting to write down phenomenological equations only with the wave-function $\psi_L$ and incoherent exciton population $N_{\rm inc}$. 
Considering the decay of $|\psi_L|^2$, we include the following phenomenological equations explicitly showing that the lower polariton wave-function converts into an incoherent exciton population due to EID $g'_{LL}$. These are written as,
\begin{eqnarray}
\hbar \dot{N}_{\rm inc}&=&-(\Gamma_x-2g'_{LL}|X|^2|\psi_L|^2)N_{\rm inc}\nonumber\\
& &+2g'_{LL}|X|^4|\psi_L|^4\\
i\hbar \dot{\psi_L}&=&\left[\tilde{\epsilon}_L+g_{LL}(|X|^2|\psi_L|^2+N_{\rm inc})\right.\nonumber\\
& &\left. -ig'_{LL}(|X|^2|\psi_L|^2+N_{\rm inc})\right]\psi_L-Cf_{\rm ext}.
\label{eq:Linc}
\end{eqnarray}
Furthermore, Eq. \ref{eq:Linc} shows that there are two contributions to the the energy shift of the lower polariton: coherent $g_{LL}|\psi_L|^2$ and incoherent $g_{LL}N_{\rm inc}$. 
The long-lived incoherent exciton population $N_{\rm inc}$ contributes to the energy blue shift of the lower polariton at large positive pump-probe delays. 
This small EID ($g'_{LL}\sim0.3g_0$) starts to appear in the lower polariton already at a cavity detuning of -0.5 meV (not shown), but further investigation is necessary to precisely determine this. 

Finally, we discuss a possible origin of these incoherent exciton generation process and EID effect. 
Since a scattering from the lower polariton state to the exciton reservoir with a large momentum does not satisfy energy-momentum conservation, we cannot employ the same scattering picture as for the upper polariton excitation case depicted in Fig. \ref{fig:scheme}. 
One possibility would be two-photon absorption process associated with a ``heating process" as in \cite{Klembt2015}, however this would not explain the detuning dependence that we observe.
A simple explanation is that there is some finite overlap of the excitonic state with the LP branch. 
While this would explain the detuning dependence, this effect is likely to be very small and depend linearly on power. 
Alternatively, the power dependence (\textit{i.e.,} EID term) of the reservoir generation could be interpreted as a polariton-polariton Auger-like recombination, which would destroy one polariton and excite another into a state which could relax into the reservoir.

\begin{figure}
\includegraphics[width=0.45\textwidth]{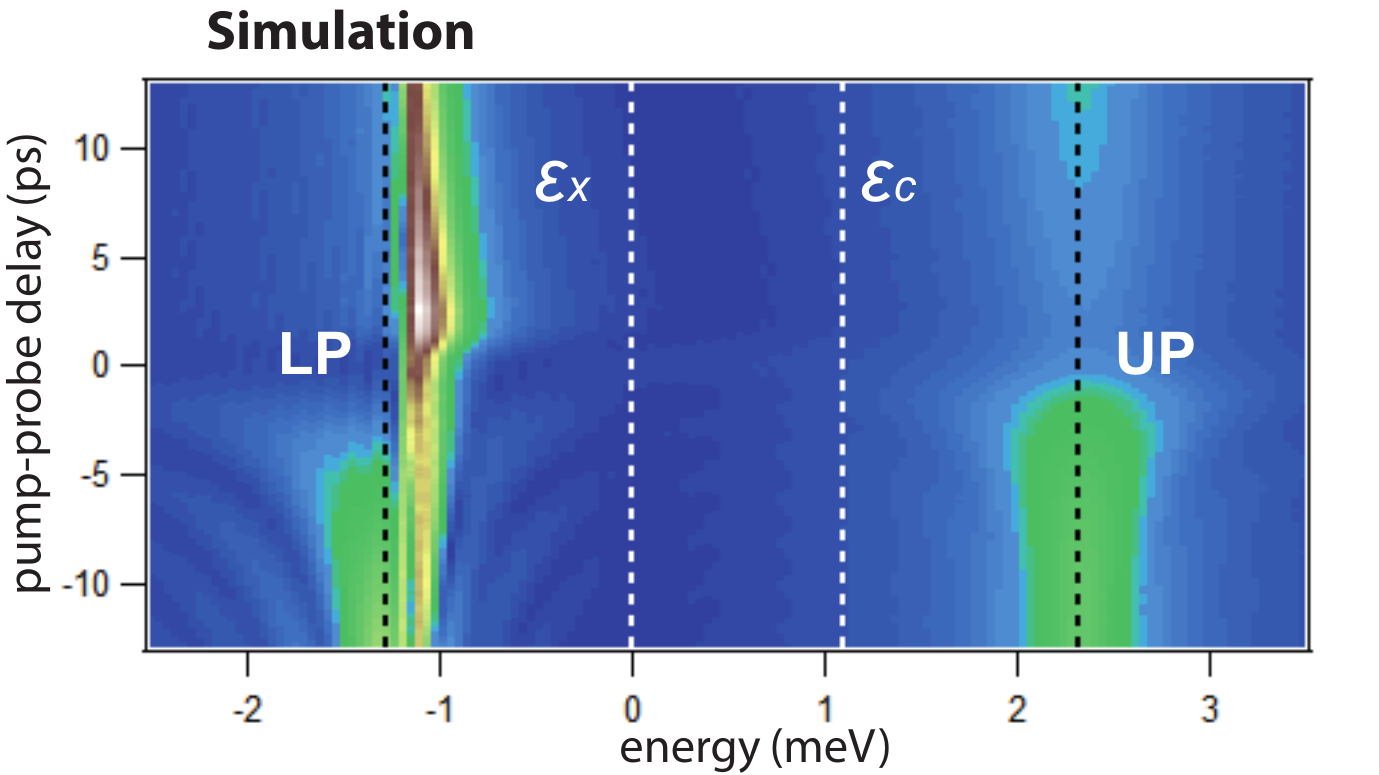}
\caption{(color online) Simulated probe transmission as a function of energy and time delay between pump and probe pulse. The additional coherent four-wave mixing term is included mixing the LP pump with the small UP population from the probe (fitting parameter: $g_{c}=10g_{LL}$)}
\label{fig:addcoh}
\end{figure}

Apart from the mechanisms described above, one should also consider a non-linear source term that is active when the polarizations induced by both the pump and probe pulses are temporally overlapping in the microcavity. 
This coherent source term corresponds to a four-wave mixing contribution that causes the scattering to the reservoir of excitons of one polariton generated by the pump pulse in the lower polariton branch with one polariton generated by the probe pulse in the upper polariton branch. 
This term takes the following form in the polariton equation of motion: \textit{i.e.,} $g_{c} \psi_{U}^{*} \psi_{L} \psi_{L}$. 
This contribution will be effective if and only if the total energy of the pump, lower polariton, and that of the probe upper polariton exceeds twice the exciton energy. 
This term was theoretically derived by Savasta \textit{et. al.,} in \cite{SavastaPRB01}; see the Eq. 7 in that article. 
By including this new coherent term in Eq. 10, we reproduce the essential features of the probe transmission signal when pumping the lower polariton branch with a spectrally narrow pulse (see Fig. \ref{fig:addcoh}): a long lived blueshifted polariton mode (at positive delay) and an enhancement of the probe transmission from negative to positive delays. 
This last feature of the data is not captured by the simulation with the standard set of polaritonic Bloch Eqs. (9),(10),(11), see comparison in Fig. \ref{fig:scheme_pos}. 
The closer agreement of the simulation presented in Fig. \ref{fig:addcoh} with the experimental data demonstrate the relevance of this new coherent term as an effective source of excitation induced dephasing, which occurs when polaritons from the pump scatter with polaritons from the probe pulse. 
This new coherent term originates from the Non-Markovian nature of the exciton-exciton interaction when going beyond the usual Hartree-Fock approximation and corresponds to four-particle correlations\cite{SavastaPRB01}.
Lastly, several state of the art microscopic calculations show that the EID strength is highly energy-dependent and becomes large when the energy of the two scattered polaritons exceeds twice the exciton energy \cite{Oestreich1995,Axt1998,Takayama2002,Kwong2001a,Savasta2003}. 
This might also explain the onset of the lower polariton EID towards positive cavity detuning. 

\section{V. Conclusion}

In conclusion, we have investigated under which conditions the polariton dynamics can be regarded as coherent or incoherent. 
For negative cavity detuning (-1.2 meV), excitation of the lower polariton can be treated in the coherent limit, thus the conventional Gross-Pitaevskii equation holds. 
On the contrary, when exciting the upper polariton branch, there is a quick transfer to the incoherent exciton population due to the strong dephasing effects for all cavity detunings investigated. 
For positive cavity detunings (\textit{e.g.,} 1.1 meV), the dephasing of lower polaritons also occurs rapidly and leads to a population in the excitonic reservoir, whatever the pumping conditions.
These findings have strong implications for the design of coherent polaritonic devices, indicating a very limited region of parameter space which is free of coherence destroying dephasing effects.     

\section*{Acknowledgement}
The present work is supported by the Swiss National Science Foundation under Project No. 153620 and the European Research Council under project Polaritonics Contract No. 291120. The polatom network is also acknowledged.



\allowdisplaybreaks[1]

\end{document}